\documentclass[12pt]{article}
\RequirePackage{graphicx}
\usepackage{amsfonts}
\usepackage{xcolor}
\usepackage{bm} 

\begin{document}
\sloppy
\vskip 1.5 truecm
\centerline{\large{\bf How to define the moving frame of the Unruh-DeWitt detector on manifolds}}
\vskip .75 truecm
\centerline{\bf Tomohiro Matsuda
\footnote{matsuda@sit.ac.jp}}
\vskip .4 truecm
\centerline {\it Laboratory of Physics, Saitama Institute of
 Technology,}
\centerline {\it Fusaiji, Okabe-machi, Saitama 369-0293, 
Japan}
\vskip 1. truecm
\makeatletter
\@addtoreset{equation}{section}
\def\theequation{\thesection.\arabic{equation}}
\makeatother
\vskip 1. truecm
\begin{abstract}
\hspace*{\parindent}
The physical phenomena seen by an observer are defined for a local
 inertial system that is subjective to the observer.
Such a coordinate system is called a ``moving frame'' because it changes
 from time to time.
However, unlike the Thomas precession, the Unruh-DeWitt detector has
 been discussed for a fixed frame.
We discuss the Unruh-DeWitt detector by defining the vacuum
for the moving frame, showing that the problem of the Stokes phenomenon
 can be solved by using the vierbeins and the exact WKB,
 to find factor
 2 discrepancy from the standard result.
Differential geometry is constructed in such a way that local
calculations can be performed rigorously.
If one expects Markov property, the calculation is expected to be local.
The final piece that was missing was a local non-perturbative
calculation, which is now complemented by the exact WKB.
Our analysis defines a serious problem regarding the relationship between entanglement of the Unruh effect and differential geometry. 
\end{abstract}

\newpage
\section{Introduction}
\hspace*{\parindent}
Physics students usually learn about local inertial systems at an
early stage in their relativity courses, and about
Thomas precession to make sense of it\cite{Misner:textbook}.
By learning Thomas precession, they learn that local inertial systems
are not only useful in curved spacetime, but are also physically
meaningful in flat spacetime.
In advanced courses, students may even learn that this local inertial 
system is related to the local trivialization of differential
geometry, and that the observer's local inertial system corresponds to a
section of the frame bundle\cite{Misner:textbook, Nakahara:textbook}.
In this way, many students learn early in their relativity courses that
the physics seen by an observer in accelerated motion must be described
by a moving frame, and how Lorentz transformation is mathematically
described.
Since the Unruh-DeWitt (UDW) detector\cite{Fulling:1972md, Davies:1974th,
Unruh:1976db} is concerned with an observer in accelerated motion, the
subjective vacuum of the observer must, in principle, be described in
terms of the moving frame. 
Calculations that can be used in such cases are commonly seen in
differential 
geometry textbooks, but they are not usually explained in papers of the
UDW detectors.\footnote{We know that there is an incorrect statement that a
moving frame is obtained by substituting the classical orbit of the
observer for the fixed frame.
The following calculations provide an easy explanation to show the error in
such a statement.} 
This shows that there is a kind of bias in papers of the
UDW detector towards introduction of a (mathematical) moving frame.
To understand what this bias is, let us first try to write down the
physical phenomena seen by the observer according to the basic
definition of the moving frame.

First define the proper time $\tau$ of the observer and denote the local
inertial system at $\tau$ by the coordinate system $X_\tau$ 
(we will explain later that $X_\tau$ is defined in the tangent space), which is
used to define the creation-annihilation operators of the observer's
subjective vacuum defined at $\tau$.
In this way, the vacuum of the observer is defined as
$|0_{M(X_\tau)}\rangle$, which must be discriminated from the
(objective) global vacuum defined for the bundle.
Here $|0_{M(X_{\tau})} \rangle$ denotes the subjective vacuum defined
for the inertial frame $X_\tau$ for the detector at its proper time $\tau$ on
the manifold $M$.
The open set on $M$ at $\tau$ is $U_{\tau}$, for which the local
trivialization is used for the observer to define the inertial frame
$X_\tau$.
As will be explained in more detail later, a clear distinction is made
here between the objective vacuum defined by the frame bundle and the
subjective vacuum defined by the section of the bundle.
The Lorenz invariance of physical quantities is explained by using the 
frame bundle.

For an observer in accelerated motion, the subjective local inertial
system ($X_\tau$) changes from moment to moment, requiring Lorentz
transformations 
(Fermi-Walker transformations) to link them.
This is called a ``moving frame''\cite{Misner:textbook}.
To avoid confusion, the following calculations follow the setups of
Ref.\cite{Birrell:textbook}. 
Using the local inertial system $X_{\tau_0}$, the Wightman Green
function at $\tau=\tau_0$ for a scalar field $\phi$ is given by
\begin{eqnarray}
\langle 0_{M(X_{\tau_0})} |\phi(x_1)\phi(x_2)|0_{M(X_{\tau_0})}\rangle
&=&\frac{-1}{4\pi^2\left[(\hat{t}-i\epsilon)^2-\hat{r}^2
\right]},
\end{eqnarray}
where $\hat{t}=t_1-t_2, {\bf \hat{r}}={\bf x_1-{\bf x_2}}$ and 
the coordinates are defined for $X_{\tau_0}$ at $\tau_0$. 
In Ref.\cite{Misner:textbook}, the range of $U_{\tau_0}$ is estimated as
$\propto a^{-1}$, where $a$ is the acceleration rate.
Strictly speaking, the above equation is valid only in
$U_{\tau_0}$, in which the local inertial frame is valid.
We now introduce the world line of the observer moving on the $z$-axis
on the coordinate system $X_{\tau_0}$;
\begin{eqnarray}
\label{eq-int0}
t(\tau)&=&a^{-1} \sinh a(\tau-\tau_0)\nonumber\\
z(\tau)&=&a^{-1} \cosh a(\tau-\tau_0),
\end{eqnarray}
which gives 
\begin{eqnarray}
\langle 0_{M(X_{\tau_0})}
 |\phi(z(\tau_0+\hat{\tau}))\phi(z(\tau_0))|0_{M(X_{\tau_0})}\rangle 
&=&\frac{-a^2}{16\pi^2\sinh^2 (a\hat{\tau}/2-i\epsilon)},
\end{eqnarray}
for which an infinite number of poles appear on the imaginary axis at
$\tau=\tau_0 (\hat{\tau}=0)$.
We will now introduce a moving frame here.
The definition of a physical quantity is an expectation value of the 
vacuum, so it has to include the coordinate system of the subjective
vacuum, because in this calculation the creation-annihilation operators
are explicitly defined for the coordinate system.
The above equation can be moved through the Lorenz transformation of the
coordinate system $X_{\tau 0}\rightarrow X_{\tau 1}$ as
\begin{eqnarray}
\langle 0_{M(X_{\tau_1})}
 |\phi(z(\tau_1+\hat{\tau}))\phi(z(\tau_1))|0_{M(X_{\tau_1})}\rangle 
&=&\frac{-a^2}{16\pi^2\sinh^2 (a\hat{\tau}/2-i\epsilon)},
\end{eqnarray}
for which an infinite number of poles appear on the imaginary axis at
$\tau=\tau_1 (\hat{\tau}=0)$.
As will be explained in more detail later, the motion
(Lorenz transformation) of the frame from $X_{\tau_0}$ to 
$X_{\tau_1}$ is defined on the frame bundle.
This result, obtained for the moving frame, clearly shows that the observer (in
constant acceleration) is seeing the same event in all local inertial
systems aligned along the trajectory, called ``moving frame''. 
On the moving frame, the observer in constant acceleration always sees
the same physical phenomena, as expected.
This result is very natural and intuitive, but makes the calculation
difficult when the vacuum response has to be obtained by integration of the
moving frame.
In the ``standard calculation'', the contribution of the poles is calculated
by extending the integration region to far outside of a given inertial
frame, where the frame is fixed and cannot be called the moving frame.
It gives
\begin{eqnarray}
&&\int^\infty_{-\infty}d\hat{\tau}e^{i \Delta E \hat{\tau}}
\langle 0_{M(X_{\tau_i})}
 |\phi(z(\tau_i+\hat{\tau}))\phi(z(\tau_i))|0_{M(X_{\tau_i})}\rangle\nonumber\\
&=&
\int^\infty_{-\infty}d\hat{\tau}e^{i \Delta E \hat{\tau}}
\sum_{n=-\infty}^{n=+\infty}\frac{-1}{4\pi^2
\left(\hat{\tau}+\frac{2\pi n}{a}i-i\epsilon\right)^2}.
\end{eqnarray}

The above calculation represents the ``standard
calculation''\cite{Birrell:textbook}. 
Note that in the ``standard calculation'' the poles 
will only appear once in the integration.
Let us now clarify the problems and the benefits of this calculation.
The ``standard calculation'' integrates the vacuum response from the
infinite past to the future {\it in the observer's proper time,
but in reality it does not consider the moving frame}.
If this integral is essential, this is implicitly a calculation for which 
the locality (Markov property) does not hold.
In practice, the response outside the valid open set is
calculated in the ``coordinate system not seen by the observer'', which
may (or, of course may not) have no effect on the result.
If the integration outside the valid open set does not
cause non-trivial effect on the calculation, one might assume that 
the local calculation on the open set $U_i$ could be equivalent to the
``standard calculation''. 
Then the calculation can be rephrased as
\begin{eqnarray}
&&\int_{U_i}d\hat{\tau}e^{i \Delta E \hat{\tau}}
\langle 0_{M(X_{\tau_i})}
 |\phi(z(\tau_i+\hat{\tau}))\phi(z(\tau_i))|0_{M(X_{\tau_i})}\rangle\nonumber\\ 
&\simeq&
\int^\infty_{-\infty}d\hat{\tau}e^{i \Delta E \hat{\tau}}
\langle 0_{M(X_{\tau_i})}
 |\phi(z(\tau_i+\hat{\tau}))\phi(z(\tau_i))|0_{M(X_{\tau_i})}\rangle,
\end{eqnarray}
where the first quantity is calculated on a given open
set\footnote{Obviously, $\tau$-integral along the moving frame 
cannot be described by just one open set.
If we stick to the definition of differential geometry, the standard
calculation is merely an extrapolation.}, while the second is the
``standard calculation''. 
Here $\int_{U_i}$ is the integration restricted to the open
set $U_i$.
The above calculation shows that the result obtained for
the ``standard calculation'' {\it could} be correct for a moving frame but it is
 neither obvious nor trivial.
Moreover, since the conventional Unruh effect anticipates entangled pair
production in distant wedges, the calculation of the detector must also
have global contribution if the results coincide.
To understand more about the local physics of the UDW detector, we searched
for the Stokes phenomena in the local inertial frame.
Since the UDW detector should always look the same physics (i.e,
the detector is expected to be always looking at the same thermal state of the
same temperature), we 
thought that the problem should be solved if the Stokes phenomenon
occurs on all local inertial systems (moving frame) aligned on the trajectory.
Since a moving frame can be described in terms of the vierbeins, it is
natural to assume that the secret lies in the vierbeins.
In this paper, by using the exact WKB and the vierbeins, we will show that the
Stokes phenomenon appears as we had expected.
This is the first time that the Stokes phenomenon of the UDW detector
has been discovered.
Interestingly, the same procedure of local trivialization that defines the
local inertial system in differential geometry also exists for gauge
transformations.
By defining the vacuum for this local system (we shall call it the local
subjective gauge), we have found in Ref.\cite{Matsuda:2023mzr, Matsuda:2025hzn} that the
Stokes phenomenon of the Schwinger effect always appears in the local system.
This approach is useful when introducing Stokes phenomena, which appear
to be dynamical, into (in a sense) static particle production such as 
the Schwinger effect, the Unruh effect and Hawking
radiation\cite{Matsuda:2025hzn}.

\subsection{A short introduction to differential geometry of the moving
 frame}
The introduction above gave a very intuitive discussion of the ``standard
calculation'' to see how it can be explained using a moving frame, but
it is probably not enough to get the full picture.
In this section we will explain in more detail how the definition of
vacuum and the introduction of the coordinate system can be explained 
in a mathematical framework. 

When the coordinate system is chosen to define the vacuum and the
 creation-annihilation operators are written, the vacuum is defined for
 this ``frame''.
The global objective vacuum can be defined collectively by using the
 frame bundle,  while the coordinate system of the subjective vacuum of
 a certain inertial observer appears to be a section of the bundle. 
As the observer accelerates, the subjective inertial frame changes with
 time.
Therefore, a subjective vacuum is defined for an observer, which is
 local for the observer, but the objective vacuum defined for the bundle is
 always global. 

\begin{figure}[ht]
\centering
\includegraphics[width=1.0\columnwidth]{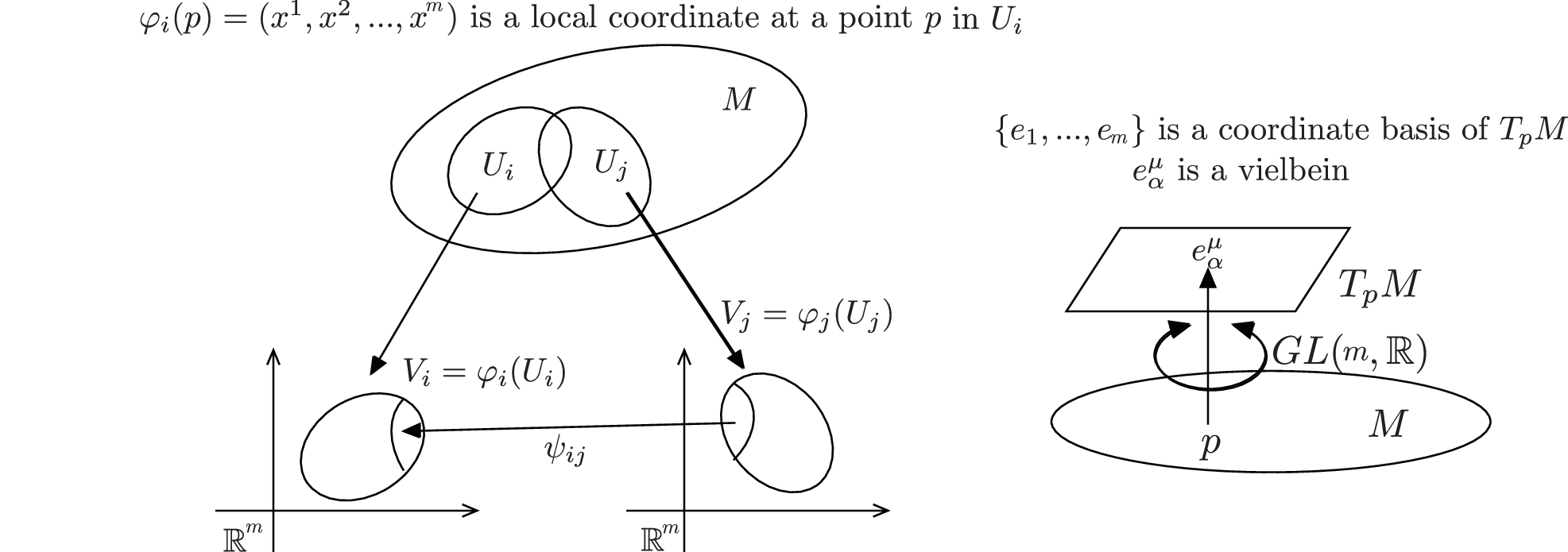}
 \caption{$m$-dimensional differentiable manifold $M$ and chatrs on $M$
 are shown in the left picture.
The tangent space and the cordinate basis, vierbeins are shown in the right.
Lorentz symmetry is explained by the vielbeins, which will be explained later.
These pictures also show why the vacuum, which respects Lorentz
 symmetry, must be defined in the tangent space.} 
\label{fig_atlas}
\end{figure}

We will first explain the intuitive reasons why a mathematical setup is
necessary.
Let us define the structure of spacetime in manifolds so that they
include relativity.
We will explain later why ``manifolds'' are plural.
We introduce $M$ as a $m$-dimensional (for our discussion we consider
$m=4$) differentiable manifold, where the question is how to introduce
coordinates so that it is compatible with relativity.
A differentiable manifold is a topological space equipped with a
structure that allows calculus to be performed {\it locally}.
It is normally defined using an atlas (a colletion of coordinate charts)
that enable differentiable transitions between local coordinate systems.
Here, a chart is a homeomorphism $\varphi_i \colon U_i\rightarrow
\mathbb{R}^m$, where $U_i$ is an open subset of the manifold $M$.
$\varphi_i$ maps points in $U_i$ to coordinates in Euclidean space, as
is shown in Fig.\ref{fig_atlas}.
Then, a coordinate basis is defined on the tangent space.
Intuitively, the freedom to choose the frame corresponds to Lorentz
symmetry.
This ``symmetry'' is introduced by the Lie group, but it is not a
trivial task to introduce the Lie brackets to the system.
We show the situation in Fig.\ref{fig_atlas}. 
The reason why local inertial frame is chosen for an observer is
understood by the procedure of local trivialisation in mathematics, but
there is an ambiguity that will be crucial for our later discussion.
For later convinience, we note that Rindler coordinates are a type of
coordinate chart used in special relativity.
In some papers there is a confusion between charts and frames, but there
is a clear distinction between the two, even in flat spacetime.

\begin{figure}[ht]
\centering
\includegraphics[width=0.8\columnwidth]{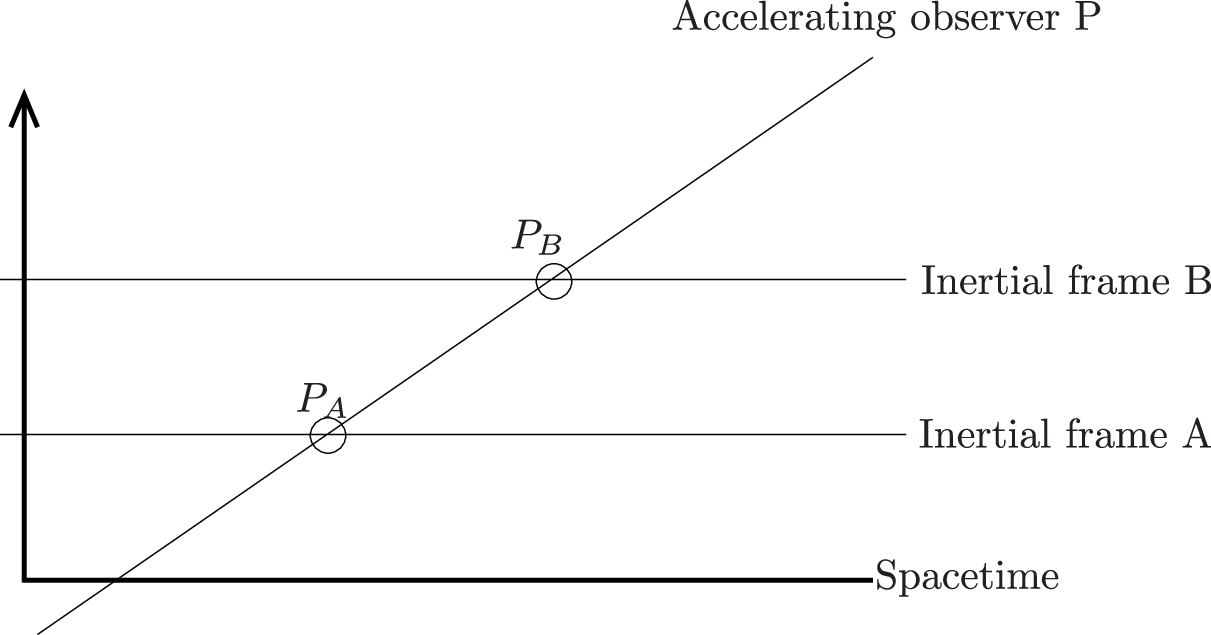}
 \caption{The vacuum seen by an accelerating observer is shown.
The observer traverses layers of local inertial systems.
The vacuum seen by the observer at $P_A$ must be defined for the
 inertial frame A, while at $P_B$ it must be defined for the inertial frame B.
The vertical axis denotes the freedom of the frame, and the horizontal axis
is the space-time.}
\label{fig_fig1}
\end{figure}

We know that the vacuum for inertial observers looks all the same at
the same time at the same place even if their inertial systems are all
different.  
This is an idea that underlies the theory of relativity, but the
situation where ``the vacuum looks exactly the same in multiple frames
all at the same time'' is difficult to imagine, especially because the
vacuum (creation-annihilation operators) is normally written down using
a specific coordinate system. 
Of course, defining it mathematically is a non-trivial task.
Currently, the vacuum as seen by any observer of any frame is defined
collectively by a ``frame bundle'', and the vacuum as seen by a specific
observer is defined as a ``section of the frame bundle'', which gives a
very clear explanation.
The problem on the physics side is that when we talk about ``vacuum'', the
discussion normally continues with ambiguity as to whether we mean 
``the vacuum defined by the observer's subjective coordinate system'' or
``the objective vacuum defined for the bundle''.
If an observer accelerates, the inertial system of the observer
changes, so the inertial system can only be defined locally.
This can be rephrased that the vacuum (written down in a specific
coordinate system) seen by an accelerated observer is defined
by the local inertial system at a given time\cite{Misner:textbook}.
The local inertial system is chosen so that the observer's velocity is
zero at that moment.
Each inertial system can be extended individually to infinity, but the
local inertial system defined for the accelerating observer
is not valid outside the neighbourhood coordinate system.
As shown in Fig.\ref{fig_fig1}, the degrees of freedom of the ``frame''
can be understood by treating them as if they were real internal spaces.
This space must be discriminated from $M$, and gives a
manifold of the frame bundle, which has the dimension $2m=8$.
Currently, the vacuum as seen by any observer of any frame is defined
collectively as a ``frame bundle'' which gives a
very clear explanation in Fig.\ref{fig_fig1}.\footnote{
Here the base space $M$ is a manifold, and the bundles and the
Lie groups are also manifolds. 
Also, later we introduce a scalar field, which is defined using a vector
space.
The vector space (and its bundle) is also a manifold.
It should be noted that the manifolds used in our discussion are
collections of different types of manifolds.}
In mathematics, such degrees of freedom are naturally treated on manifolds.
It is important to note that in mathematics there is no need for
``observers'' in the description of the space-time structure (described
by the manifolds), while
physics always uses an observer, so we normally see a
``section'' of the mathematically described manifold (bundle).
The reason for the difficulty of this story is that the equations can
only be defined locally if the observer traverses the frames.\footnote{A
similar problem was widely recognized first in 
the Dirac monopole solution\cite{Dirac:1931kp}:
when constructing the Dirac monopole solution, the equation must be
solved on at least two open sets\cite{Wu:1975es}, and the two
solutions are laminated by using the gauge transformation.}
For example, for observers in the accelerating system in
Fig.\ref{fig_fig1}, the vacuum 
equation defined by the inertial system A is only correct in the
vicinity of $P_A$.
If one wants to integrate the effect of the vacuum felt by the observer
using the frame coordinates in a mathematically correct way, one has to laminate
open sets in which the local inertial systems are defined.
Indeed, there are examples, such as Thomas precession, where it is very
important to analyze the acceleration system in the observer-specific
frame, where lamination by the Lorentz transformation is essential.
To construct a whole from what can only be defined locally, knowledge of
differential geometry can be used.\footnote{The textbook by Misner et
al.\cite{Misner:textbook} provides a detailed description of the vacuum
 seen by an accelerating observer in section 6. 
The textbook also explains the Thomas precession in detail.
However, the mathematical ideas in defining such a vacuum, such as fibre and
bundle, are not discussed explicitly in the text.
Ref.\cite{Nakahara:textbook} will complement the missing parts.} 
As we have discussed above, the vacuum of the UDW 
detector\cite{Fulling:1972md, Davies:1974th, Unruh:1976db} has to be
discussed in terms of local inertial systems, not (at least in
principle) for the coordinates of a global (or fixed) frame.
When calculating the UDW detector, it could be plausible that fixing the
frame of the vacuum using a global coordinate system may not destroy the
essential physics of the Unruh effect. 
On the other hand, we already know that the observer's frame cannot be fixed in Thomas
precession calculations. 
Therefore, it is still not obvious if one can use the coordinates of the
local inertial system outside the neighbourhood
to integrate the response of the UDW detector\cite{Birrell:textbook}.
Our conclusion in this paper will be that extrapolating the local frame to
infinity leads to a factor 2 discrepancy in the non-perturbative factor.
The non-perturbative factor corresponds to the Boltzmann factor.
Given that differential geometry and manifolds make the computation
local, it is clear that the crucial question was whether a local method of
computation in such a framework existed. 
The standard calculation of the UDW detector is global, but such a
computation cannot be justified in the above framework.
Out point is that local calculation of the non-perturbative effect became
possible by the mathematical framework of the exact WKB.

More formal mathematics required for our argument is summarised below.
To avoid confusion, our mathematical definitions follow Nakahara's
textbook\cite{Nakahara:textbook} as far as possible.

The base space $M$ is an
$m$-dimensional differentiable manifold \footnote{This ``M'' is not for Minkovski} with a 
family of pairs (called chart\footnote{The Rindler coordinates give a
coordinate chart representing part of flat
Minkowski space-time.}) $\{(U_i,\varphi_i)\}$, where $\{U_i\}$ is a
family of open sets (coordinate neighbourhoods) which covers $M$ as $\bigcup_i
U_i=M$, and $\varphi_i$ (coordinate function\footnote{In our
introduction we have used $X_{\tau}$ for the inertial frame, which must
be discriminated from this coordinate function. More details will be
explained later in this section.} $\{x^1(p),...,x^m(p)\}$ at
$p\in U_i$) is a homeomorphism from $U_i$ onto an open subset of
$\mathbb{R}^m$.
A tangent bundle $T M$ over $M$ is a collection of all tangent spaces of
$M$: 
\begin{eqnarray}
TM\equiv \bigcup_{p\in M}T_p M.
\end{eqnarray}
Suppose that $\varphi_i(p)$ is the coordinate function $\{x^\mu (p)\}$.
Note that the mathematical argument becomes trivial if the base space is
flat, but the section of the fibre that the observer cuts is
non-trivial\footnote{The term ``non-trivial'' here needs to be
distinguished from the mathematical term ``trivial''.}
in the following arguments.
In the ``coordinate basis'', $T_p M$ is spanned by
$\{e_\mu\}=\{\partial/\partial x^\mu\}$, while the ``non-coordinate
bases'' is explained as
\begin{eqnarray}
\hat{e}_\alpha&=&e_\alpha^\mu\frac{\partial}{\partial x^\mu},\,\,\,\,
e_\alpha^\mu\in GL(m,\mathbb{R}),
\end{eqnarray}
where the coefficients $e_\alpha^\mu$ are called vierbeins (or more
generally called vielbeins if it is many dimensional).
{\it Obviously, the Lie bracket can be introduced only for the non-coordinate bases.}
There exists a dual vector space to $T_p M$, whose element is a linear
function from $T_pM$ to $\mathbb{R}$.
The dual space is called the cotangent space at $p$, denoted by $T^*_p M$.
Since $U_i$ is homeomorphic to an open subset $\varphi(U_i)$ of
$\mathbb{R}^m$ and each $T_pM$ is homeomorphic to $\mathbb{R}^m$,
$TU_i\equiv \bigcup_{p\in U_i}T_pM$ is a $2m$-dimensional manifold,
which can be (always) decomposed into a direct product $U_i\times
\mathbb{R}^m$.
Given a principal fibre bundle $P(M,G)$, one can define an associated
fibre bundle as follows.
For $G$ (a group) acting on a manifold $F$ on the left, one can define
an action 
of $g\in G$ on $P\times F$ by
\begin{eqnarray}
(u,f)&\rightarrow&(ug,g^{-1} f)
\end{eqnarray}
where $u\in P$ and $f\in F$.
Now the associated fibre bundle is an equivalence class $P\times F/G$ in
which $(u,f)$ and $(ug,g^{-1} f)$ are identified.

For a point $u\in TU_i$, one can systematically decompose the
information of $u$ into $p\in M$ and $V\in T_pM$.
This leads to the projection $\pi$ : $TU_i\rightarrow U_i$, for which
the information about the vector $V$ is completely lost.
Inversely, $\pi^{-1}(p)=T_pM$ is what is called the fibre at $p$. 

Normally, one requires that $\hat{e}_\alpha$ be orthonormal with respect
to g;
\begin{eqnarray}
\mathrm{g}(\hat{e}_\alpha,\hat{e}_\beta)=e_\alpha^\mu e_\beta^\nu
 \mathrm{g}_{\mu\nu}=\delta_{\alpha\beta},
\end{eqnarray}
where $\delta_{\alpha\beta}$ must be replaced by $\eta_{\alpha\beta}$ for
the Lorentzian manifold.
The metric is obtained by reversing the equation
\begin{eqnarray}
 \mathrm{g}_{\mu\nu}=e^\alpha_\mu e^\beta_\nu\delta_{\alpha\beta}.
\end{eqnarray}
In an $m$-dimensional Riemannian manifold, the metric tensor
$\mathrm{g}_{\mu\nu}$ has $m(m+1)/2$ degrees of freedom while the vielbein
has $m^2$ degrees of freedom.
Each of the bases can be related to the other by the local orthogonal
rotation $SO(m)$, while for Lorentzian manifold it becomes $SO(m-1,1)$.
The dimension of these Lie groups is given by the difference between the degrees
of freedom of the vielbein and the metric.
Therefore, there are many non-coordinate bases that yield the identical
metric.
This point will be very important when one looks at the Unruh
effect\cite{Fulling:1972md, Davies:1974th, Unruh:1976db}. 
The local inertial frame and the Lorentz frame have the same metric and
are sometimes used as if they are interchangeable and define the same
vacuum, but they must be distinguished by the vierbein.
The difference is crucial when covariant derivatives are defined
since the vierbein must be diagonalized (i.e, twists and rotations must
be removed) to define the covariant derivatives.
This indicates that the field equation (i.e, covariant derivatives) on a
curved space-time may not see (at least directly) the ``inertial vacuum''.
The difference is important in the search for Stokes phenomena in the
Unruh effect\cite{Matsuda:2023mzr}.
In this paper, we are focusing on the Stokes phenomenon of the
UDW detector, which is defined on a flat space-time.

For a more formal explanation
 of the meaning of Fig.\ref{fig_fig1},
we describe the frame bundle below.
Associated with a tangent bundle $TM$ over $M$ is a principal bundle
called the frame bundle $LM\equiv \bigcup_{p\in M}L_p M$
where $L_p M$ is the set of frames at $p$.
Since the bundle $T_p M$ has a natural coordinate basis 
$\{\partial/\partial x^\mu\}$ on $U_i$, a ``frame''
 $u=\{X_1,...,X_m\}$ at $p$ is expressed by the non-coordinate basis
\begin{eqnarray}
X_\alpha&=&X^\mu_\alpha \left.\frac{\partial}{\partial x^\mu}\right|_p
\end{eqnarray}
 where $(X^\mu_\alpha)\in GL(m,\mathbb{R})$.   
If $\{X_\alpha\}$ is normalized by introducing a metric, the matrix
$(X_\alpha^\mu)$ becomes the vielbein.
Then the local trivialization is
$\phi_i$ : $U_i \times GL(m,\mathbb{R})$ $\rightarrow$ 
$\pi^{-1}(U_i)$ by $\phi^{-1}_i(u)=(p,(X^\mu_\alpha))$.

\begin{figure}[ht]
\centering
\includegraphics[width=1.0\columnwidth]{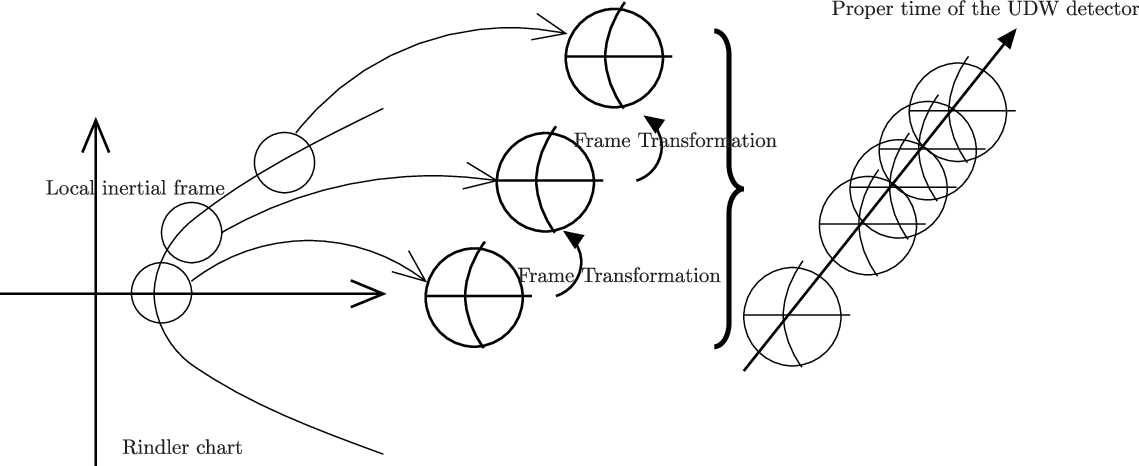}
 \caption{The inertial frame on the Rindler coordinate chart is explained.
For observers in uniformly accelerated motion, the situation around them
 always looks the same.
To illustrate the situation, after considering the local inertial system
 in each open set, the coordinates of the inertial system are
 transformed into Cartesian coordinates and taken out to the right.
It can be seen that the use of just one of the local inertial systems as
 a whole system is a rather wild approximation.
The validity of such procedures must be confirmed by local analysis.}
\label{fig_movingframe}
\end{figure}

Now that the mathematics is ready, we will have a look at the
content of Fig.\ref{fig_fig1} with the help of the mathematics.
Typically, a natural coordinate basis is prepared on the surface of $M$
and the inertial system is defined using a non-coordinate basis.
This procedure naturally gives a ``moving frame'' explained in
Ref.\cite{Misner:textbook}.
See also our Fig.\ref{fig_movingframe}.

\begin{figure}[ht]
\centering
\includegraphics[width=0.8\columnwidth]{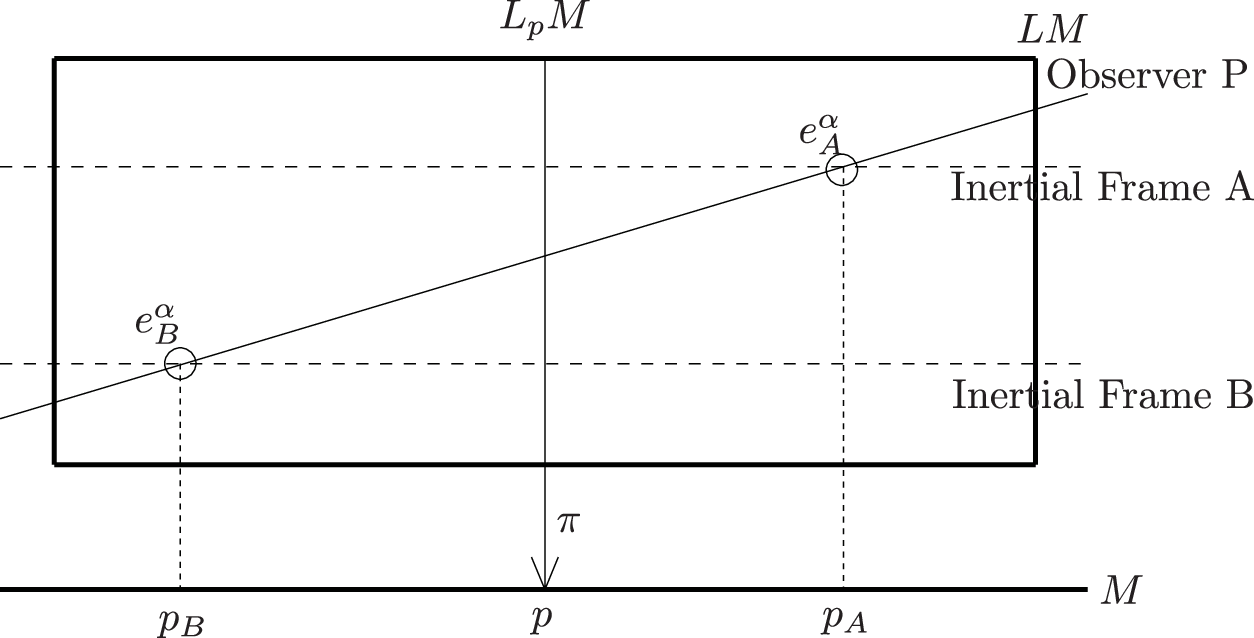}
 \caption{The situation is shown for the frame bundle $LM$.
The mathematical description does not require observers, while physics
 requires an observer, which defines a section of the bundle.
The small circles are written on the frame bundle for illustration but
 should be written on $M$ as they correspond to $U_i$.}
\label{fig_fig2}
\end{figure}

However, since Fig.\ref{fig_fig1} describes an observer in accelerated
motion on a flat space-time, one tends to define a ``global (fixed) inertial
system'' on the Cartesian chart and define an accelerating observer on it.
This is the reverse sequence of the standard mathematical procedure and
introduces serious confusion about the nature of the space-time structure.
If one continues with such a  (wrong) definition, we
believe the definition of the 
frame bundle discussed above cannot be used naively, and
the ``moving frame'' cannot be described.
Therefore, to avoid confusion, we first define the inertial system using
a non-coordinate basis, as we have done for defining the frame bundle.
In this case, the inertial system A is defined on the tangent space at
$P_A$ in Fig.\ref{fig_fig1}, and the inertial system B is defined on the
tangent space at $P_B$ using the vielbeins $(e_A)_\alpha^\mu$ and
$(e_B)_\alpha^\mu$, respectively.
To be more specific, the vielbeins for constant acceleration
(denoted by $a$ in the following) in the two-dimensional space-time can be described as
\begin{eqnarray}
(e_A)_\alpha^\mu&=&\left(
\begin{array}{cc}
\cosh a(\tau -\tau_A)&\sinh a(\tau -\tau_A)\\
\sinh a(\tau -\tau_A)&\cosh a(\tau -\tau_A)
\end{array}
\right)\\
(e_B)_\alpha^\mu&=&\left(
\begin{array}{cc}
\cosh a(\tau -\tau_B)&\sinh a(\tau -\tau_B)\\
\sinh a(\tau -\tau_B)&\cosh a(\tau -\tau_B)
\end{array}
\right),
\end{eqnarray}
and the transformation between them is 
\begin{eqnarray}
L_{AB}&=&\left(
\begin{array}{cc}
\cosh a(\tau_B -\tau_A)&-\sinh a(\tau_B -\tau_A)\\
-\sinh a(\tau_B -\tau_A)&\cosh a(\tau_B -\tau_A)
\end{array}
\right), 
\end{eqnarray}
for which we have $ e_A\,  L_{AB}=e_B$.
These relations and the definition of the
vacuum are obvious in terms of the frame bundle.
As is shown in Fig.\ref{fig_fig2}, the detector (observer) in physics
determines its local inertial frame.
Each inertial system can be extended individually to infinity, but the local inertial systems defined for the observer of the accelerating system are not all the same, so there is an effective range for the local inertial systems.
In the Thomas precession calculations, it was essential that these
inertial systems should be laminated by Lorentz transformations.
The mathematical description of the frame bundle
does not require an observer, while physics inevitably introduces an
observer, which defines a section of the bundle.
Although the mathematical description of the frame bundle in the flat
space-time is quite trivial and only one chart is enough,
in our introduction the chart has been segmented to describe
the local inertial vacuum (the moving frame). 
The above vielbeins (for the inertial frame) are not for the
mathematical Lorentz frame because they have a twist.
Unlike the inertial frame, the Lorentz frame is normally
defined using a diagonal vielbein. 
As the covariant derivatives are defined for the Lorentz frame, 
this makes it difficult to examine the Stokes phenomena of the Unruh
effect directly in terms of the field equations\cite{Matsuda:2023mzr},
as far as the twist in the inertial frame is essential for the UDW
detector.
More precisely, both inertial and Lorentz frames are
diagonal at a point (because $\sinh 0=0$), but the inertial frame is
twisted in the neighbourhood.
Note that the Stokes phenomenon of the Unruh effect is very different
from the Schwinger effect\cite{Schwinger:1951nm} in the sense that 
the Stokes phenomenon of the Schwinger effect can be obtained very easily 
from the field equations.
In contrast to the simplicity of the Stokes phenomenon, the mathematical
structure of the Schwinger effect is more complex than that of the Unruh
effect in the sense that the Schwinger effect requires both the frame bundle
and the gauge bundle at the same time\cite{Matsuda:2023mzr}.
(The Schwinger effect requires the ``moving frame'' for particles and the
``moving gauge'' at the same time.)

Next, we mention semi-classical approximation and singular
perturbations in quantum mechanics as a preparation for further
mathematics to describe the Stokes phenomenon.
In most textbooks of quantum mechanics, it is said that the limit of
$\hbar\rightarrow 0$ is a ``semi-classical'' limit, but in reality,
the story has to be more complicated.
Since the equations of quantum mechanics have $\hbar$ 
in the coefficients of the derivatives, $\hbar \rightarrow 0$ 
is the ``singular perturbation'' where the rank of the differential
equation changes.
Very naively, if the wave function (solutions of the field equation) is given by
analytic functions $\psi(z)=e^{S(z)}$, 
 the mapping could be discontinuous at $\Im[S]=2\pi n$, where $n$ is 
an integer.
When one studies the Stokes phenomenon of such solutions, one has to
study the behaviour of the solutions at such discontinuity
line\footnote{This line is called the Stokes line,
where two $\pm$ solutions mix. As a very complex process is required to
explain why two solutions given by regular functions mix, we would like
the reader to consider that the mixing occurs because they are
discontinuous there. See Ref.\cite{ExactWKB:textbook, Virtual:2015HKT}
for more mathematical details.}, paying 
attention to singular perturbations. 
Technically, a technique called resurgence\cite{RPN:2017, Voros:1983,
Delabaere:1993, Silverstone:2008} is used in this analysis.
The underlying ideology will not resemble the so-called semi-classical
``approximation''.
The series of $\hbar$-expansion is analytically continued to the complex
$\eta$-plane ($\eta\equiv \hbar^{-1}$), and the divergent power series
of the WKB expansion ($\eta^{-n}$-expansion) is transformed into a
finite integral by Borel summation.
The argument is primarily based on analytic continuation
rather than asymptotic expansion.
The Borel summation maps the functions of $\eta$ into the functions on
the Borel panel.
We thus have a complex variable {\it in addition to } the conventional
coordinate.
Investigation shows that the singularity at the
turning point\footnote{Since we are dealing here with the ``Schr\"odinger
equation'' used by mathematicians, the turning point is nothing but 
the turning point used in ordinary quantum mechanics.} is related to the
discontinuity on the Stokes lines and the Stokes phenomenon. 
The basic form of the current exact WKB was mostly designed by Pham et. al in
Refs.\cite{Pham:1988, CNP:1993, DDP:1993,DDP:1997} and then extended by
many people including the group of the authors of the
textbooks\cite{ExactWKB:textbook, Virtual:2015HKT}. 

Although ``WKB'' is used in the name, the actual analysis of the exact WKB is
more concerned with singular perturbations than with the semi-classical
approximation.
In this paper, the exact WKB is used for the Stokes phenomena,
following the textbook\cite{ExactWKB:textbook} by Kawai and Takei and 
Ref.\cite{Virtual:2015HKT} by Honda, Kawai and Takei.
We will not go into the details of mathematics and will use only the
fruitful results of the exact WKB, which should be supplemented by these
textbooks. 
We believe that those who may not be convinced of the local description of
the Stokes phenomenon will find these textbooks very useful.
In addition, if one might question the special role of
$\eta=\hbar^{-1}$, it would be helpful to read the references
regarding that $\hbar\rightarrow 0$ is a singular perturbation.
In that respect, $\hbar$ is already special.

This paper is organized as follows.
In section \ref{sec-Howto}, we describe how to define the UDW 
detector on manifolds, using the basic idea of the frame bundle.
Then, in section \ref{sec-EWKB}, we explain how to find the Stokes
phenomenon of the UDW detector in an open set.

\section{How to define the Unruh-DeWitt detector on manifolds}
\hspace*{\parindent}
\label{sec-Howto}
Let us first take a look at the calculations of the Unruh-DeWitt
detector\cite{Davies:1974th, Unruh:1976db} in the
textbook\cite{Birrell:textbook} by Birrell and Davies  
and then examine it using the frame bundle\cite{Nakahara:textbook}.
It is not possible to cover everything here. 
For other calculations and approaches please refer to the review
paper\cite{Crispino:2007eb}.

\begin{figure}[ht]
\centering
\includegraphics[width=1.0\columnwidth]{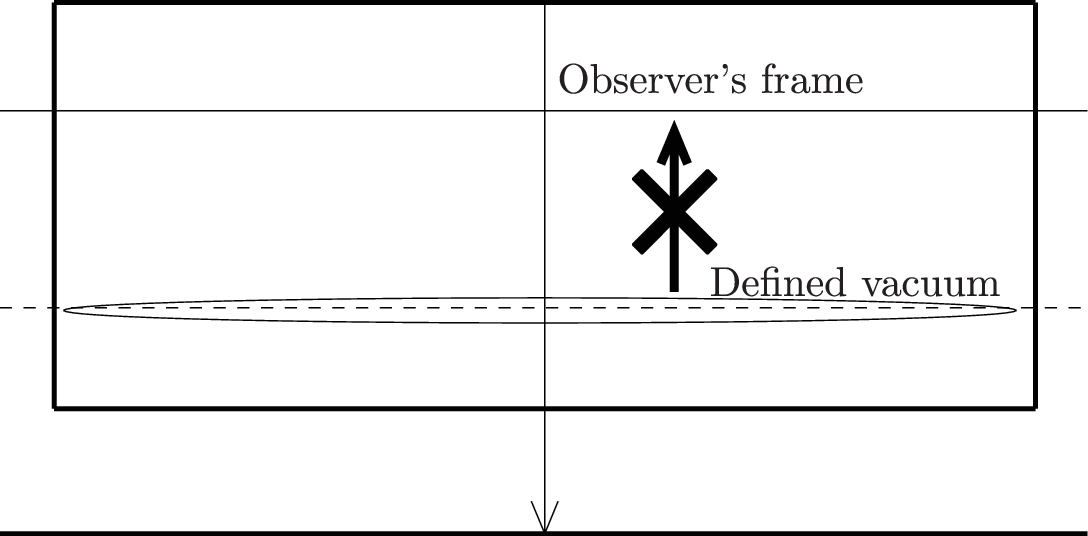}
 \caption{The definition of the vacuum in Eq.(3.51) of
 Ref.\cite{Birrell:textbook} is misleading. 
As is shown in this figure, the vacuum must be defined for the
 observer's frame, which must not have velocity if it is seen by the
 observer.} 
\label{fig_constv}
\end{figure}

Following Ref.\cite{Birrell:textbook}, let us introduce a particle
detector that moves along the worldline described by the functions
$x^\mu(\tau)$, where $\tau$ is the detector's proper time.
The detector-field interaction is described by 
${\cal L}_\mathrm{int}= c\,\, m(\tau) \phi[x^\mu(\tau)]$, where $c$ is a small
coupling constant and $m$ is the detector's operator.
Suppose that the scalar field $\phi$ describes the vacuum state as
\begin{eqnarray}
\phi(t,{\bm{\mathrm{x}}})&=&\sum_{\bm{\mathrm{k}}} 
\left[
a_{\bm{\mathrm{k}}} u_{\bm{\mathrm{k}}}(t,{\bm{\mathrm{x}}})
+a_{\bm{\mathrm{k}}}^\dagger u_{\bm{\mathrm{k}}}^*(t,{\bm{\mathrm{x}}})
\right]\nonumber\\
a_{\bm{\mathrm{k}}} |0\rangle&=&0.
\end{eqnarray}
We assume that $\sum_{\bm{\mathrm{k}}}$ can be replaced by integration.
For sufficiently small $c$, the amplitude from $|0\rangle$ to an excited
state $|\psi\rangle$ may be given by the first-order perturbation as
\begin{eqnarray}
\label{eq-int}
ic\langle E,\psi|\int^\infty_{-\infty}m(\tau)\phi[x^\mu(\tau)]d\tau|0,E_0\rangle.
\end{eqnarray}
Using the Hamiltonian $H_0$, one has 
$m(\tau)=e^{i H_0 \tau}m(0)e^{-iH_0 \tau}$,
where $H_0 |E\rangle=E|E\rangle$.
One can factorize the amplitude as
\begin{eqnarray}
\label{eq-amplitude}
ic\langle E|m(0)|E_0\rangle \int^\infty_{-\infty}e^{(E-E_0)\tau}
\langle\psi|\phi(x)|0\rangle d \tau.
\end{eqnarray}
As we are considering only the first-order transition, the excited state
is the state $|\psi\rangle=|1_{\bm{\mathrm{k}}}\rangle$, which contains
only one quantum.
Then, one has
\begin{eqnarray}
\langle\psi|\phi(x)|0\rangle&=&\langle
 1_{\bm{\mathrm{k}}}|\phi(x)|0\rangle\\
&=&\int d^3k' (16\pi^3 \omega')^{-1/2}
\langle 1_{\bm{\mathrm{k}}}|a_{\bm{\mathrm{k'}}}^\dagger|0\rangle 
e^{-i {\bm{\mathrm{k}}}'\cdot {\bm{\mathrm{x}}}+i\omega' t}.
\end{eqnarray}
For an inertial detector, an ``inertial worldline'' is introduced in
Ref.\cite{Birrell:textbook} as
\begin{eqnarray}
\label{eq-inerline}
{\bm{\mathrm{x}}}&=&{\bm{\mathrm{x_0}}}+{\bm{\mathrm{v}}}t\\
&=&{\bm{\mathrm{x_0}}}+{\bm{\mathrm{v}}}\tau(1-v^2)^{-1/2},
\end{eqnarray}
where $v$ is a constant velocity.
It is claimed that 
\begin{eqnarray}
\label{eq-claim}
&&(16\pi^3 \omega)^{-1/2}e^{-i {\bm{\mathrm{k}}}\cdot {\bm{\mathrm{x_0}}}}
\int^\infty_{-\infty} e^{i(E-E_0)\tau}
e^{i\tau(\omega-  {\bm{\mathrm{k}}}\cdot {\bm{\mathrm{v}}})
(1-v^2)^{-1/2}}d\tau\nonumber\\
&=&(4\pi \omega)^{1/2}e^{-i {\bm{\mathrm{k}}}\cdot {\bm{\mathrm{x_0}}}}
\delta(E-E_0+(\omega-  {\bm{\mathrm{k}}}\cdot {\bm{\mathrm{v}}})
(1-v^2)^{-1/2}).
\end{eqnarray}
This vanishes because $(E-E_0+(\omega- {\bm{\mathrm{k}}}\cdot
{\bm{\mathrm{v}}})(1-v^2)^{-1/2})> 0$.
Surprisingly, in the above calculation taken from
Ref.\cite{Birrell:textbook}, the vacuum is {\it not} defined for the subjective
inertial frame of the detector. 
We show this situation in Fig.\ref{fig_constv}.
This calculation is quite {\it misleading} even if the correct result
could be obtained in this way.
Without acceleration, the frame does not move and the Lorentz
transition is trivial for this calculation.
In such cases, using a different frame may not cause major problems.
However, as is shown in Fig.\ref{fig_constv}, the vacuum defined for a distant
frame cannot be projected to the observer's true vacuum by using the ``inertial
worldline'' of Eq.(\ref{eq-inerline}).
Many might think it is pointless to bother with such trivial
calculation in the textbook, but in Ref.\cite{Birrell:textbook} it is
explained later that by making this  
worldline (or trajectory) more complex, the excitation can be seen in
this formalism, and in fact, many papers citing this textbook actually
define the vacuum using the same concept and perform similar calculations. 
Therefore, this is a problem that cannot be overlooked.
The correct calculation is that the detector has 
\begin{eqnarray}
\langle
 1_{\bm{\mathrm{k}}}|\phi(x)|0\rangle
&=&\int d^3k' (16\pi^3 \omega')^{-1/2}
\langle 1_{\bm{\mathrm{k}}}|a_{\bm{\mathrm{k'}}}^\dagger|0\rangle 
e^{-i {\bm{\mathrm{k}}}'\cdot {\bm{\mathrm{x}}}+i\omega' \tau}
\end{eqnarray}
for the detector's inertial coordinates $({\bm{\mathrm{x}}},\tau)$.
This simply gives instead of Eq.(\ref{eq-claim}),
\begin{eqnarray}
(16\pi^3 \omega)^{-1/2}e^{-i {\bm{\mathrm{k}}}\cdot {\bm{\mathrm{x}}}}
\int^\infty_{-\infty} e^{i(E-E_0)\tau}
e^{i\tau \omega}d\tau.
\end{eqnarray}
The amplitude of transition must vanish because $E-E_0+\omega>0$.
The point is that one cannot introduce $v\ne 0$ (a relative velocity between the
vacuum coordinates and the observer)  when the vacuum is defined
(or chosen) properly for the observer's subjective frame.
For an accelerating observer, this point can be rephrased that one
cannot introduce $v\ne 0$ for the local inertial frame in the $U_i$ on $M$.
The confusing point would be that in the textbook\cite{Birrell:textbook} ``the rest frame of the
moving detector'' is not identical to ``the frame of the vacuum'' as is
depicted in Fig.\ref{fig_constv}.
As far as a non-accelerating system is considered, it
is easy to arrive at the correct answer even with this kind of
treatment.
However, such a treatment causes great confusion in the analysis of
acceleration systems, as it blurs the concept of moving frame and local
inertial systems.

\begin{figure}[ht]
\centering
\includegraphics[width=1.0\columnwidth]{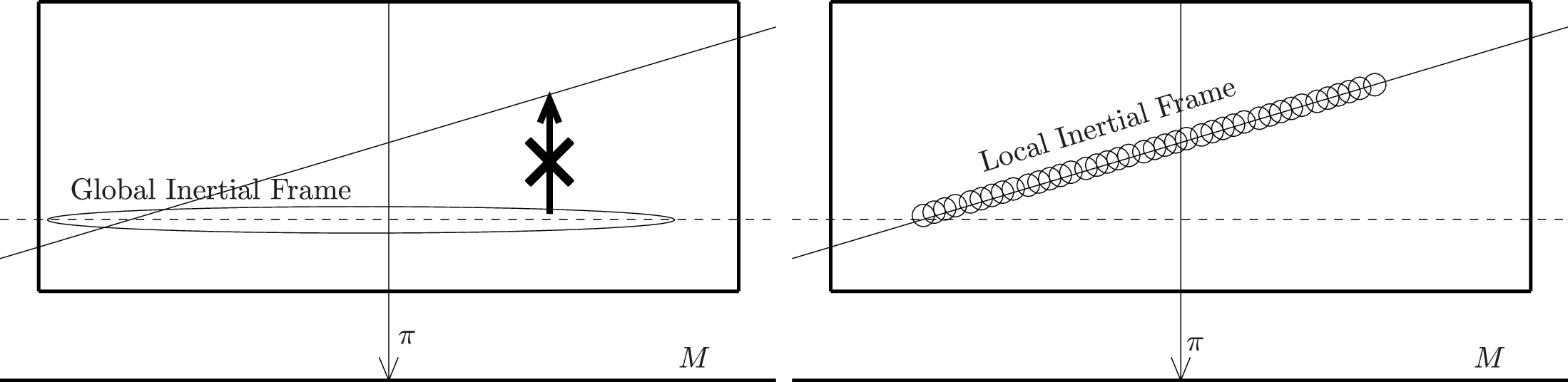}
 \caption{These figures show how to define the frames for the vacuum.
On the left panel, the vacuum is defined for a global inertial
 frame, which is a section of the frame bundle.
On the right panel, the vacuum is defined for the local inertial frame
 in the open set $U_i$, where the vierbein is defined locally.
The small circles on the frame bundle correspond to
 the open sets  ($U_i$) on $M$.
The local inertial frame in the right panel is what is called a ``moving
 frame'' in  Ref.\cite{Misner:textbook}.} 
\label{fig_accel}
\end{figure}

Next, paying careful attention to the above calculation, let us consider
the case where the detector is moving in a constant accelerating motion.
In this case, the (subjective) vacuum can only be defined locally as the observer
traverses the frame bundle, as is shown in the right panel of Fig.\ref{fig_accel}.
In differential geometry, the integral in such cases is often written as
follows;
\begin{eqnarray}
\int_M \hat{\omega} = \sum_i \int_{U_i} \rho_i \hat{\omega},
\end{eqnarray}
where $\hat{\omega}$ is normally a $m$-form and $\rho_i$ is required for
lamination.
In our case, the integral with respect to the proper time $\tau$ must be
segmented since the vacuum is defined only locally.\footnote{We believe
discrimination between the (subjective) local vacuum and the (objective)
global vacuum is already very clear in this paper.}
When performing calculations such as those in the left panel of
Fig.\ref{fig_accel}, it might be predicted that non-trivial phenomena
will only be seen at the intersection, since the only place where the
vacuum is actually seen is at the intersection of the two lines.
Indeed, the calculation in Ref.\cite{Birrell:textbook} shows poles on
the imaginary axis only at that point.
Even if the calculation shows symmetry regarding time translations,
integrating over the distant vacuum is not justified.
It should be noted that the calculation of the pole contribution\cite{Birrell:textbook}
considers integration on the other parts of the vacuum where the distant
vacuum is defined.

Noting that the vierbein connects the inertial frame (vacuum) and the
observer, we rewrite $\phi$ as
\begin{eqnarray}
\phi(t,{\bm{\mathrm{x}}})&=&\sum_{\bm{\mathrm{k}}} e^{-i {\bm{\mathrm{k}}}\cdot
{\bm{\mathrm{x}}}}\left[
a_{\bm{\mathrm{k}}} e^{-i\int \omega dt}
+a_{-\bm{\mathrm{k}}}^\dagger e^{+i\int \omega dt}
\right],
\end{eqnarray}
where $\omega$ is constant but $dt$ is required to introduce
vierbein in an explicit form.
Obviously, the transition amplitude of Eq.(\ref{eq-amplitude}) 
does not vanish if $e^{\pm i\int \omega dt}$ are mixed since
the amplitude for $E-E_0-\omega =0$ is possible after the mixing.
We expect that the mixing can be observed when they are seen by the detector.
To be more specific, if a vierbein $(e(\tau))^t_\tau$ gives $dt=
(e(\tau))^t_\tau d\tau$, mixing
of the 
solutions after crossing the Stokes line can be written as
\begin{eqnarray}
e^{\pm i\omega \int^\tau   (e(\tau'))^t_\tau d\tau'}&\rightarrow&\alpha_\pm
e^{\pm i\omega\int^\tau  (e(\tau'))^t_\tau d\tau'} 
+\beta_\pm e^{\mp i\omega \int^\tau
(e(\tau'))^t_\tau d\tau'}.
\end{eqnarray}
Now we have 
\begin{eqnarray}
\phi(t,{\bm{\mathrm{x}}})&=&\sum_{\bm{\mathrm{k}}} 
e^{-i {\bm{\mathrm{k}}}\cdot
{\bm{\mathrm{x}}}}\left[
a_{\bm{\mathrm{k}}} 
\left(\alpha_- e^{- i\omega\int^\tau  (e(\tau'))^t_\tau d\tau'} 
+\beta_- e^{+ i\omega \int^\tau
(e(\tau'))^t_\tau d\tau'}
\right)
\right.\nonumber\\
&&\left.+a_{-\bm{\mathrm{k}}}^\dagger 
\left(
\alpha_+
e^{i\omega\int^\tau  (e(\tau'))^t_\tau d\tau'} 
+\beta_+ e^{- i\omega \int^\tau
(e(\tau'))^t_\tau d\tau'}\right)
\right],
\end{eqnarray}
which suggests that the amplitude is proportional to
$\beta_+$.\footnote{For fermions, we have  
\begin{eqnarray}
\partial_\mu \psi&=&e^\alpha_\mu \partial_\alpha \psi.
\end{eqnarray}
Focusing on the time-dependent component, it can be seen that the same
function is obtained as for the scalar field.}

The $\tau$-integration in Eq.(\ref{eq-amplitude})
must be considered carefully
since it must be segmented and requires lamination.
We will be back to this issue later after explaining the Stokes phenomenon.
Now it is obvious that the Stokes phenomenon of
function $e^{\pm i\omega \int^\tau   (e(\tau'))^t_\tau d\tau'}$ needs to be investigated.
In this case, the Stokes phenomenon is not thought to change
the definition of the vacuum, but rather to make the coupled field
($\phi$) of the detector appears to be mixed when it is seen by the
accelerating detector.
To show that such a Stokes phenomenon does occur for an accelerating
observer, an analysis using the exact WKB will be presented in the following
section. 

\section{The Stokes phenomenon of the Unruh-DeWitt detector}
\label{sec-EWKB}
\hspace*{\parindent}
Typically, the exact WKB uses $\eta\equiv \hbar^{-1}\gg 1$, instead of the
Planck constant $\hbar$.
Following Refs.\cite{ExactWKB:textbook,Virtual:2015HKT}, our starting
point is the second-order ordinary differential equation given
by
\begin{eqnarray}
\left[-\frac{d^2}{dx^2}+\eta^2 Q(x,\eta)
\right]\psi(x,\eta)&=&0,
\end{eqnarray}
where both $x$ and $\eta$ will be considered as complex.
This equation is called the ``Schr\"odinger equation'' by mathematicians.
If the solution $\psi$ is written as $\psi(x,\eta)=e^{R(x,\eta)}$,
we have 
\begin{eqnarray}
\psi&=&e^{\int^x_{x_0}S(x',\eta)dx'}
\end{eqnarray}
for $S(x,\eta)\equiv \partial R/\partial x$.
Just for simplicity of the argument, we choose
$Q(x,\eta)=Q(x)$.\footnote{Although our later discussion uses higher 
terms of $Q(x,\eta)$, we are confined here in $Q(x,\eta)=Q(x)$ because the
extension is straightforward\cite{Koike:2000}.}
For $S(x,\eta)$, we have 
\begin{eqnarray}
-\left(S^2 +\frac{\partial S}{\partial x}\right)+\eta^2 Q&=&0.
\end{eqnarray}
If one expands $S$ as $S(x,\eta)=\sum_{n=-1}^{n=\infty}\eta^{-n} S_{n}$,
one will find
\begin{eqnarray}
S=\eta S_{-1}(x)+ S_0(x)+\eta^{-1}S_1(x)+...,
\end{eqnarray}
which leads to
\begin{eqnarray}
S_{-1}^2&=&Q\\
2S_{-1}S_j&=&-\left[\sum_{k+l=j-1,k\ge 0,l\ge 0}S_kS_l + \frac{d
	       S_{j-1}}{dx}\right]\\
&&(j\ge 0).\nonumber
\end{eqnarray}
The above calculation is nothing but the conventional WKB expansion.
Note however that since $\eta$ will be analytically continued, the
expansion considered here is not only for small real $\eta$ but will be
considered on the complex $\eta$-plane.

Also, note that the divergent power series are considered for a function
on the complex $\eta$-plane.
We have complex $\eta$ and $x$ at the same time, and the Borel summation
will be calculated for $\eta$.

Two power series solutions $S^{\pm}(x,\eta)$ are obtained according to
the signs of the initial term $S_{-1}=\pm \sqrt{Q(x)}$.
After defining $S_{odd}$ and $S_{even}$ by
\begin{eqnarray}
\label{eq-prove2}
S^{\pm}=\pm S_{odd}+S_{even},
\end{eqnarray}
and 
using the relation between $S_{odd}$ and $S_{even}$
\begin{eqnarray}
S_{even}&=&-\frac{1}{2}\log S_{odd},
\end{eqnarray}
 one will have 
\begin{eqnarray}
\psi&=&\frac{1}{\sqrt{S_{odd}}}e^{\int^x_{x_0}S_{odd}(x')dx'}\\
&&S_{odd}\equiv\sum_{j\ge 0}\eta^{1-2j}S_{2j-1}.
\end{eqnarray}
Depending on the sign of the first $S_{-1}=\pm \sqrt{Q(x)}$, there are
two solutions $\psi_\pm$, which are given by a simple form
\begin{eqnarray}
\psi_{\pm}&=&\frac{1}{\sqrt{S_{odd}}}\exp\left(\pm \int^x_{x_0}S_{odd}(x') dx'\right).
\end{eqnarray}
As far as there is no discontinuity (the Stokes line) in a domain,
these solutions are not mixed.
The domain is called the Stokes domain.

\begin{figure}[t]
\centering
\includegraphics[width=0.8\columnwidth]{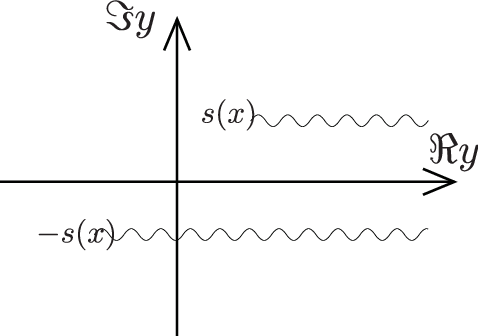}
 \caption{The wavy lines are the paths of integration of the Borel
 summation.
The complex $y$-plane is called the Borel panel.
One can see that the paths overlap when $\Im[s]=0$.}
\label{fig_borelintegral}
\end{figure}

The above WKB expansion is usually divergent but is Borel-summable.
Namely, one may consider 
\begin{eqnarray}
\psi_\pm &\rightarrow&\Psi_\pm\equiv\int^\infty_{\mp s(x)}e^{-y\eta}\psi_\pm^B(x,y)dy,\\
&&s(x)\equiv \int^x_{x_0}S_{-1}(x')dx',
\end{eqnarray}
where the $y$-integral is parallel to the real axis.\footnote{Normally,
one can choose the integration on the steepest descent path.}
The Borel summation can be considered as the conventional Laplace
transformation back from $\psi^B$, and $\psi^B$ is obtained by the Borel
transformation (almost equivalent to the inverse Laplace transformation)
 of the original function.
Therefore, one can see that the original function is transformed as
$\psi_\pm\rightarrow\psi^B_\pm\rightarrow\Psi_\pm$, where the final
function corresponds to the original function 
written using the Borel summation.
The easiest way of explaining the Stokes phenomenon is to use the Airy function
($Q(x)=x$) near the turning points.
What is important here is that motion on the complex $x$-plane
causes $s(x)$ to move on the complex $y$-plane.
If one defines the Stokes line starting from the turning point
at $x=0$ as 
\begin{eqnarray}
\Im [s(x)]=0.
\end{eqnarray}
The Stokes lines are the solutions of
\begin{eqnarray}
\Im [s(x)]&=& \Im \left[\int^x_0 (x')^{1/2}dx' \right]\nonumber\\
&=&\Im \left[\frac{2}{3}x^{3/2}\right]=0,
\end{eqnarray}
which can be written as ``three straight lines coming out of the turning
point placed at the origin''.
The paths of integration on the $y$-plane, which explains the Stokes
phenomenon when the end-points ($\pm s(x)$) cross the integration
contour,  are shown in Fig.\ref{fig_borelintegral}.
Note that the integration paths overlap on the Stokes line since the Stokes lines
are defined as the solutions of $\Im [s(x)]=0$.
From Fig.\ref{fig_borelintegral}, one can understand why additional
contributions (mixing of the solutions) can appear after crossing the
Stokes line at $\Im [s(x)]=0$.
This ``reconnection of the integration path in the Borel
plane'' causes mixing of the $\pm$ solutions and is called the Stokes
phenomenon. 

The usual WKB expansion can also be used to discuss Stokes lines, but it
cannot be said that the Stokes lines are strictly described by $s(x)$ alone.
The usual WKB approximation cannot mention whether higher order terms in
$\hbar$ change the shape of the Stokes line.
This point is crucial for our analysis.
Although physics involves various forms of approximation and expansion,
it can be said that any approximation that completely changes the nature
of the Stokes line is not an appropriate approximation.
In Refs.\cite{Enomoto:2022asc,Enomoto:2021alz}, we found that such
``improper approximation'' does exist in past studies, and showed
explicitly that such approximation can spoil discussion of cosmological
particle-antiparticle asymmetry. 
We stress that when dealing with a complex Stokes phenomenon, the original
Stokes lines should be checked before considering approximations.
In the discussion here, it was possible that even if the Stokes lines
were written, the Stokes phenomenon might not have occurred in the
neighbourhood.
As will be discussed later, the fact that the exact Stokes line runs
through the neighbourhood allowed the proper approximation to be
performed.

For the consideration of the Stokes phenomenon of the Unruh effect, we
make use of a very characteristic property of the exact WKB:
as we have described above, the Stokes lines are determined by 
$\Im [s(x)]=0$, where $s(x)$ is defined only by using $S_{-1}$.
This is not the result of a mere approximation, but the result of
considering singular perturbations.
If a higher-order term were important in the determination of
the Stokes line, analyses using the Stokes line would always have run
the risk that corrections by the higher-order terms would alter the
fundamental properties obtained from the lower-order terms.
Specific examples can be found in Ref.\cite{Enomoto:2021alz}.
In the following, we will investigate the Stokes phenomenon of the
``solutions in the inertial system when they are seen by an
accelerating observer''.
Such an analysis would not have been possible without the characteristic
properties of the exact WKB, for the reasons we have described above
and below.

Let us see how the Stokes phenomenon of the UDW detector
appears.
In the open set $U_i$ defined for $\tau=\tau_i$, 
an accelerating observer is looking at the
local inertial vacuum using the local vierbein
$(e_i)^\mu_\alpha$.
Here we consider only the time-dependent part and use 
$dt=\cosh a(\tau-\tau_i) d\tau$ with $\tau_i=0$
to rewrite the local inertial vacuum solutions.
We also introduce explicit $\eta$ as
\begin{eqnarray}
\label{eq-rindler-sol}
e^{\pm i \int^t \omega dt'}&\rightarrow&e^{\pm i \eta \int^\tau \omega \cosh(a \tau') d\tau'}.
\end{eqnarray}
As we have already discussed, it should be sufficient for us now to
examine the nature of this solution.
However, normally, it is quite difficult to recognize the Stokes phenomenon of
 such solutions.
Our idea is that using the characteristic properties of the exact WKB mentioned
above, one can reconstruct the Stokes phenomenon.
Let us first introduce $Q_0(\tau)\equiv -\omega^2\cosh^2(a \tau)$ and
consider 
 the following equation
\begin{eqnarray}
\label{eq-rindler-EoM}
\left(-\frac{d^2}{d\tau^2}+\eta^2 Q(\tau,\eta)\right)\psi(\tau,\eta)&=&0,
\end{eqnarray}
where $\eta\gg 1$ and $Q(\tau,\eta)$ is expanded as
\begin{eqnarray}
Q(\tau,\eta)&=&Q_0(\tau)+\eta^{-1}Q_1(\tau)+\eta^{-2}Q(\tau)+\cdots.
\end{eqnarray}
Note that, unlike the normal procedure, the terms other than $Q_0$ have
not yet been determined.
As we have done before, the solution of this equation can be written as $\psi(\tau,\eta)\equiv
e^{\int^\tau S(\tau',\eta) d\tau'}$, where $S(\tau,\eta)$ can be expanded as 
\begin{eqnarray}
S&=&S_{-1}(\tau)\eta +S_0(\tau)+S_1(\tau)\eta^{-1}+\cdots.
\end{eqnarray}
The point of the above argument is that after introducing $\eta$ 
in Eq.(\ref{eq-rindler-sol}), one can choose $Q_i(\tau)$ $(i=1,2,...)$ to
reconstruct the equation that gives the solutions
Eq.(\ref{eq-rindler-sol}). 
Here $Q_i$ $(i\ge 1)$ has to be chosen so that $S_i$ $(i\ne
-1)$ in $S_{odd}$ vanishes in the final solution (\ref{eq-rindler-sol}).
Details of the expansion are shown below to make the calculation easier
to imagine.
Again, we start with the solution
\begin{eqnarray}
\psi&=&e^{\int^\tau_{\tau_0}S(\tau',\eta)d\tau'},
\end{eqnarray}
where $S$ is determined by the equation
\begin{eqnarray}
\label{eq-prove1}
-\left(S^2 +\frac{\partial S}{\partial \tau}\right)+\eta^2 Q&=&0.
\end{eqnarray}
As $S$ is expanded as 
\begin{eqnarray}
S=\eta S_{-1}(\tau)+ S_0(\tau)+\eta^{-1}S_1(\tau)+...,
\end{eqnarray}
and $Q(\tau,\eta)$ is expanded as
\begin{eqnarray}
Q(\tau,\eta)&=&Q_0(\tau)+\eta^{-1}Q_1(\tau)+\eta^{-2}Q(\tau)+\cdots,
\end{eqnarray}
one will find 
\begin{eqnarray}
S_{-1}^2&=&Q_0\nonumber\\
2S_{-1}S_0+ \frac{dS_{-1}}{d\tau}&=&Q_1(\tau)\nonumber\\
2S_{-1}S_1+S_0^2+ \frac{dS_{0}}{d\tau}&=&Q_2(\tau)\nonumber\\
&...&,
\end{eqnarray}
where we demand $S_{2j-1}=0$ for $j>0$ to determine $Q(\tau,\eta)$.
Note also that we still have 
\begin{eqnarray}
S_{even}&=&-\frac{1}{2}\log S_{odd},
\end{eqnarray}
which can be obtained from Eq.(\ref{eq-prove2}) and (\ref{eq-prove1}).

\begin{figure}[ht]
\centering
\includegraphics[width=0.8\columnwidth]{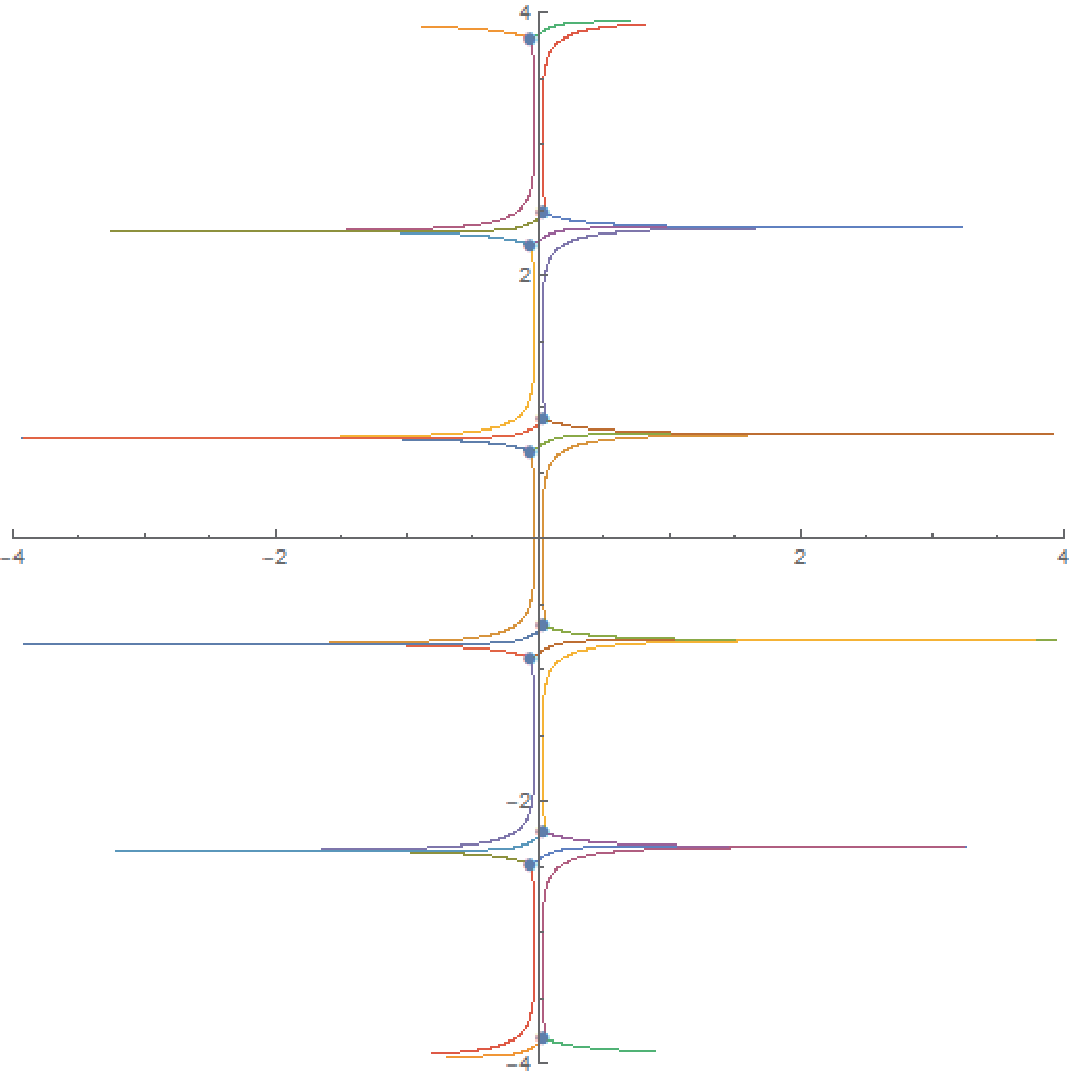}
 \caption{The Stokes lines of the Unruh effect are shown for
 $Q(\tau)=-\left(\cosh ^2(2 \tau)+0.05 i-0.05\right)(1-0.05 i) $. The degenerated Stokes
 lines are separated introducing small parameters.}
\label{fig-UdW-stokes}
\end{figure}

As we have mentioned above, this procedure allows us to make use of a
powerful analysis of the exact WKB.
After drawing the Stokes lines, one can see that a Stokes line crosses on the
real axis at the origin\cite{Enomoto:2022mti}.
The Stokes lines of the Unruh effect are shown in
Fig.\ref{fig-UdW-stokes}.
One can easily check that the Stokes lines cross the origin.
This allowed us an approximation: expand $Q(\tau)_0$ near the origin
without changing the crossing point.
The approximation gives 
\begin{eqnarray}
\label{eq-app}
Q(\tau)_0&=& -\omega^2\cosh(a \tau)\nonumber\\
&\simeq&-\omega^2-a^2 \omega^2 \tau^2,
\end{eqnarray}
whose Stokes lines are the same as the familiar Schr\"odinger equation of scattering by an
inverted quadratic potential.
The equation can be solved using the parabolic cylinder functions or the
Weber functions, giving a very characteristic structure of the Stokes
lines\cite{Enomoto:2022mti, Enomoto:2020xlf}.\footnote{To be more
precise, there is no confirmation that such special functions solve the
equation when the higher terms $Q_i(\tau),(i\ge 1)$ are introduced in
the equation.
The main advantage of the exact WKB is that it shows that the structure
of the Stokes line remains unchanged for such ``perturbation''.
If the Stokes lines are unchanged, the difference appears
only in the higher terms in the integral between the turning points,
whose contribution can be disregarded for smal $\hbar$.}

The Stokes phenomenon occurs at $\tau=0$ in the above calculation, which
corresponds to the point where the local inertial vacuum is defined.
(The Stokes phenomenon occurs at $\tau=\tau_i$ for $U_i$ defined for
$\tau_i$.)
Therefore, the same Stokes phenomenon can be seen in each $U_i$.
Laminating such $U_i$ on $M$ using the frame transformation, one will
find that stationary (continuous) excitation should be observed by the detector.

Now we can define the UDW  detector on manifolds.
Our starting point is the detector defined on the local inertial frame of an
open set $U_i$.
For sufficiently small $c$ in Eq.(\ref{eq-int}), the
amplitude from 
$|0\rangle$ to an excited 
state $|\psi\rangle$ may be given by the first-order perturbation as
\begin{eqnarray}
ic\langle E,\psi|\int_{U_i}m(t)\phi d\tau|0,E_0\rangle.
\end{eqnarray}
Then, using the Hamiltonian $H_0$, one has 
$m(t)=e^{i H_0 t}m(0)e^{-iH_0 t}$, where $H_0 |E\rangle=E|E\rangle$.
These quantities are defined for the local inertial frame
on $U_i$.
One can factorize the amplitude as
\begin{eqnarray}
ic\langle E|m(0)|E_0\rangle \int_{U_i}e^{(E-E_0)t}
\langle\psi|\phi|0\rangle d \tau,
\end{eqnarray}
where $t$ is used because $H_0$ is originally defined for the inertial
frame.
We are discriminating $t$ in the tangent space and $\tau$ for the detector.
As we are considering only the first-order transition, the excited state
is the state $|\psi\rangle=|1_{\bm{\mathrm{k}}}\rangle$, which contains
only one quantum.
Then, one has\footnote{
The Fourier transformation is considered in the
tangent space where the vacuum is defined. 
The calculation here follows Sec.\ref{sec-Howto}.}
\begin{eqnarray}
\langle\psi|\phi(x)|0\rangle&=&\langle
 1_{\bm{\mathrm{k}}}|\phi(x)|0\rangle\\
&=&\int d^3k' (16\pi^3 \omega')^{-1/2}
\langle 1_{\bm{\mathrm{k}}}|a_{\bm{\mathrm{k}}'}^\dagger|0\rangle 
e^{-i {\bm{\mathrm{k}}'}\cdot {\bm{\mathrm{x}}}+i\int^t \omega' dt'}.
\end{eqnarray}
As we have described above, $e^{i\int^t \omega dt'}$ experiences the
Stokes phenomena on each $U_i$ when it is seen by an accelerating
observer.
Using Eq.(\ref{eq-app}) and the standard calculation of
the Schr\"odinger equation for scattering by an inverted potential, 
the Stokes phenomenon is described by the connection 
matrix\cite{Enomoto:2020xlf};
\begin{eqnarray}
\left(
\begin{array}{cc}
\alpha_+ & \beta_+\\
\beta_- & \alpha_- \\
\end{array}
\right)
&=&
\left(
\begin{array}{cc}
\sqrt{1+e^{-2K_{ud}}} & -i e^{-K_{ud}} \\
 i e^{-K_{ud}} & \sqrt{1+e^{-2K_{ud}}} \\
\end{array}
\right),
\end{eqnarray}
where the integration factor $K_{ud}=\int^u_d
S_{odd} d\tau'$ is the integration connecting the 
two turning points (i.e, two solutions of $Q_0=0$) on the imaginary
axis.
The phases are omitted for simplicity.
Here the solution is written according to the manner of
the exact WKB, but readers who are familiar with special functions may
directly use special functions to find the answer.
These two turning points are calculated after the approximation of
Eq.(\ref{eq-app}).
Finally, we obtain the amplitude of the transition caused by the Stokes
phenomenon in $U_i$.
The standard calculation of the scattering problem by an inverted
quadratic potential gives $|\beta_+|^2\simeq e^{-\frac{\omega}{a}\pi}$,
which corresponds to the Boltzmann factor.
The most important aspect of this result is that the Boltzmann factor
differs by a factor of 2 from the usual UDW detector calculation
($\sim e^{-\frac{2\pi\omega}{a}}$).
Indeed, if the standard calculation of the UDW detector includes the
production of entangled 
particles at a distant wedge, the result should differ from our local
calculation by a factor of 2, because the particles at that distant
wedge are not detected and the probability of detecting ``a'' particle
is given by a pair production $\sim (e^{-\frac{\omega}{a}\pi})^2$.
The difference arose because in our calculations everything is local,
faithful to the definition of differential geometry and the
Markov property.
Ultimately, the question can be divided into two parts.
One is the question of whether entanglement of the Unruh effect is
realistic.
Our position is that the entanglement is not real because we believe that
the local analysis by differential geometry is correct.
We interpret this as the result of extending local systems
outside the applicable range, which makes it impossible to maintain
mathematical consistency.
The other is the possibility that local descriptions of the above
differential geometry cannot be applied to quantum entanglements.
In this direction, there may be a need for an extension such as
Penrose's twistor 
theory\cite{Penrose:1967twistor}, but the question is still open. 
It should also be noted that no such discussion has been made for the
standard Unruh effect calculations before.

The problem discussed here does not arise in Hawking radiation, because
in Hawking radiation a pair of particles is produced and only one of the
particles is observed as radiation.
Pair production in Hawking radiation occurs locally on the horizon, so
there is no need to consider the problem of distant wedges.
The construction of the field theory by differential geometry described
here does not present a problem in Hawking
radiation\cite{Matsuda:2025hzn}.

\section{Conclusions and Discussions}
\label{sec-concdis}
\hspace*{\parindent}
In this paper, we have described how to define the UDW 
detector for a moving frame.
In defining the vacuum and its coordinate system, we were particularly
careful not to introduce 
relative velocity with the observer.
Since the local inertial frame can only be defined locally as the observer
accelerates, a locally defined Stokes phenomenon was inevitable.
The situation seems to be similar to the monopole solution.
The first solution (the Dirac monopole solution) was given by Dirac in
1931, but a singularity remained until Wu and Yang solved it using
differential geometry in 1975.
For the Unruh effect, the recent development of the exact WKB was inevitable
for finding the local Stokes phenomenon on the local inertial frame.
To say nothing of the Dirac monopole as an example, the usual
construction of field theory can be elaborated and given many facets by
means of differential geometry.

Mathematically, it is not an obvious approximation to use the local
inertial system outside the neighbourhood coordinate system to deal with
integrals with respect to the observer's time. 
This leads to the fact that the poles only appear in the
neighbourhood when integrating the Green's function, and there is (in
principle) room for improvement of the calculation. 
We have shown that the UDW detector can be treated locally as defined in
differential geometry, without extrapolating the local inertial system
and extending it to infinity. 
By using local analysis, our work establishes the computation of the UDW
detector in terms of the differential geometry.

The most important result of our calculation is that the Boltzmann
factor differs by a factor of 2 from the usual calculation of the UDW
detector and the Unruh effect.
Since the standard calculation of the Unruh effect includes the
production of a pair of entangled particles at distant wedges, 
it is natural to find that our result differs from the standard
calculation by a factor of 2.
This factor arises because the particles at that distant
wedge are not detected by the detector and only the probability changes
for a pair production.
The crucial difference arose because in our calculation everything was local,
faithful to the concept of differential geometry.
Here the question can be divided into two parts.
One is whether entanglement of the Unruh effect is realistic or not, and 
the other is the possibility that the standard description of the field
theory by means of the differential geometry could be wrong for 
 quantum entanglements.

The problem discussed here does not arise in Hawking radiation, since
in Hawking radiation a pair of particles is produced locally at the
horizon and also only one of the particles is observed as radiation.
All events can be calculated locally using the conventional differential
geometry and there is no factor 2 problem.

We hope that the new perspectives presented in this paper will help us
to understand and improve the physics of the Unruh effect and the
Unruh-DeWitt detector.

\section{Acknowledgment}
\hspace*{\parindent}
The author would like to thank all the participants and the organisers
of the Third Hermann Minkowski Meeting on the Foundations of Spacetime
Physics in 2023 for giving him the opportunity to have very useful
discussions.
In particular, he is deeply grateful to Professor Bertolami for his
helpful comment on the Unruh-DeWitt detector.
The author is also thankful to Yasutaka Takanishi for carefully reading
the manuscript and providing useful advice.


\begin{thebibliography}{1}
\bibitem{Misner:textbook}
C.~W.~Misner, K.~S.~Thorne and J.~A.~Wheeler,
``Gravitation,''
W. H. Freeman, 1973,
ISBN 978-0-7167-0344-0, 978-0-691-17779-3.
\bibitem{Nakahara:textbook}
M.~Nakahara, 
``Geometry, Topology and Physics,''
CRC Press, 2003, 
ISBN 978-0-7503-0606-5, 978-0-7503-0606-5.
\bibitem{Fulling:1972md}
S.~A.~Fulling,
``Nonuniqueness of canonical field quantization in Riemannian space-time,''
Phys. Rev. D \textbf{7} (1973), 2850-2862.
\bibitem{Davies:1974th}
P.~C.~W.~Davies,
``Scalar particle production in Schwarzschild and Rindler metrics,''
J. Phys. A \textbf{8} (1975), 609-616.
\bibitem{Unruh:1976db}
W.~G.~Unruh,
``Notes on black hole evaporation,''
Phys. Rev. D \textbf{14} (1976), 870.
\bibitem{Birrell:textbook}
N.~D.~Birrell and P.~C.~W.~Davies, ``Quantum Fields in Curved Space,''
Cambridge University Press, 1984, 978-0-5212-7858-4.
\bibitem{Matsuda:2023mzr}
T.~Matsuda,
``Nonperturbative particle production and differential geometry,''
Int. J. Mod. Phys. A \textbf{38} (2023) no.28, 2350158,
[arXiv:2303.11521 [hep-th]].
\bibitem{Dirac:1931kp}
P.~A.~M.~Dirac,
``Quantised singularities in the electromagnetic field,,''
Proc. Roy. Soc. Lond. A \textbf{133} (1931) no.821, 60-72.
\bibitem{Wu:1975es}
T.~T.~Wu and C.~N.~Yang,
``Concept of Nonintegrable Phase Factors and Global Formulation of Gauge Fields,''
Phys. Rev. D \textbf{12} (1975), 3845-3857.
\bibitem{Schwinger:1951nm}
J.~S.~Schwinger,
``On gauge invariance and vacuum polarization,''
Phys. Rev. \textbf{82} (1951), 664-679.
\bibitem{RPN:2017}
``Resurgence, Physics and Numbers'' edited by F.~Fauvet, D.~Manchon,
	S.~Marmi and  D.~Sauzin, Publications of the Scuola Normale
	Superiore, 978-88-7642-613-1.
\bibitem{Voros:1983}
A.~Voros, ``The return of the quartic oscillator -- The complex WKB
method'', Ann. Inst. Henri Poincare, 39 (1983), 211-338.
\bibitem{Delabaere:1993}
E. Delabaere, H. Dillinger and F. Pham: Resurgence de Voros et peeriodes
des courves hyperelliptique. Annales de l'Institut Fourier, 43 (1993), 163-
199.
\bibitem{Silverstone:2008}
H.~Shen and H.~J.~Silverstone, ``Observations on the JWKB treatment of
the quadratic barrier, Algebraic analysis of differential equations from
	microlocal analysis to exponential asymptotics'', Springer,
	2008, pp. 237 - 250.
\bibitem{Pham:1988}
F.~Pham, 
``Resurgence, quantized canonical transformations, and multiinstanton
expansions,''
Algebraic Analysis, Vol. II, Academie Press, 1988, pp. 699-726.
\bibitem{CNP:1993}
B.~Candelpergher, J.~G.~Nositicls and F.~Pham,
``Approche de la Resurgence,''
Hermann, 1993.
\bibitem{DDP:1993}
E.~Delabaere, H.~Dillinger and F.~Pham, 
``Resurgence de Voros et periodes des courbes hyper elliptiques,''
 Ann. Inst. Fourier (Grenoble) 4 3 (1993), 163-199.
\bibitem{DDP:1997}
E.~Delabaere, H.~Dillinger and F.~Pham,
``Exact semi-classical expansions for one dimensional quantum oscillators,''
J. Math. Phys. 38 (1997), 6126-6184.
\bibitem{ExactWKB:textbook}
T.~Kawai and Y.~Takei,
``Algebraic Analysis of Singular Perturbation Theory,''
Iwanami Series in Modern Mathematics, 2005, 978-0-8218-3547-0.
\bibitem{Virtual:2015HKT}
N.~Honda, T.~Kawai and Y.~Takei,
``Virtual Turning Points'', Springer (2015),  978-4-431-55702-9.
\bibitem{Crispino:2007eb}
L.~C.~B.~Crispino, A.~Higuchi and G.~E.~A.~Matsas,
``The Unruh effect and its applications,''
Rev. Mod. Phys. \textbf{80} (2008), 787-838,
[arXiv:0710.5373 [gr-qc]].
\bibitem{Enomoto:2022asc}
S.~Enomoto and T.~Matsuda,
``The Exact WKB analysis for asymmetric scalar preheating,''
JHEP \textbf{01} (2023), 088,
[arXiv:2203.04497 [hep-th]].
\bibitem{Enomoto:2021alz}
S.~Enomoto and T.~Matsuda,
``The exact WKB and the Landau-Zener transition for asymmetry in cosmological particle production,''
JHEP \textbf{02} (2022), 131,
[arXiv:2104.02312 [hep-th]].
\bibitem{Enomoto:2022mti}
S.~Enomoto and T.~Matsuda,
``The Exact WKB analysis and the Stokes phenomena of the Unruh effect and Hawking radiation,''
JHEP \textbf{12} (2022), 037,
[arXiv:2203.04501 [hep-th]].
\bibitem{Enomoto:2020xlf}
S.~Enomoto and T.~Matsuda,
``The exact WKB for cosmological particle production,''
JHEP \textbf{03} (2021), 090,
[arXiv:2010.14835 [hep-ph]].
\bibitem{Matsuda:2025hzn}
T.~Matsuda,
``Quantum field theory on curved manifolds,''
[arXiv:2501.09919 [hep-th]].
\bibitem{Koike:2000}
T~Koike, 
``Asymtotics of the spectrum of Heun's equation and the exact WKB
	analysis'', Towards the Exact WKB analysis of Differential
	Equations, Linear or Non-Linear (Kyoto, 1998), Kyoto
	Univ. Press(2000),55-70
\bibitem{Penrose:1967twistor}
R.~Penrose, J. Math. Phys. 8, 345 (1967).
\end{thebibliography}
\end{document}